\documentclass[proceedings%
]{aofa}

\usepackage[utf8]{inputenc}
\usepackage{subfigure}

\author[J. Gaither and M.~D. Ward]{Jeffrey Gaither\addressmark{1}\and
  Mark Daniel Ward\addressmark{2}\thanks{M.~D. Ward's work is
    supported by the Center for Science of Information (CSoI), an NSF
    Science and Technology Center, under grant agreement CCF-0939370;
    his work is also supported by NSF DMS-124681 and NSF DMS-1560332.}}

\title{Variance of the Internal Profile in Suffix Trees}

\address{\addressmark{1}Mathematical Biosciences Institute of The Ohio
  State University,
Columbus, OH 43210, USA\\
\addressmark{2}Department of Statistics, Purdue University,
West Lafayette, IN 47907, USA}
\usepackage{comment}
\usepackage{amsmath}
\usepackage{multicol}

\usepackage{subfigure}
\usepackage{epstopdf}
\usepackage{dsfont}

\newtheorem{theorem}{Theorem}
\newtheorem{lemma}{Lemma}
\newtheorem{corollary}{Corollary}
\newtheorem{proposition}{Proposition}

\newcommand{\Tst}{\textbf{T}_{\sigma,\theta}}

\def\Var{\mathop{\operatorname{Var}}}
\def\Cov{\mathop{\operatorname{Cov}}}

\newcommand{\Iu}{I_{n,u}}
\newcommand{\Iv}{I_{n,v}}

\newcommand{\gt}{g_{w,\sigma,\theta}}

\newcommand{\sumu}{\sum_{u\in\Ak}}

\newcommand{\Ak}{\mathcal{A}^{k}}

\newcommand{\sumuv}{\sum_{\substack{u,v\in\Ak \\ u \neq v}}}

\newcommand{\Puv}{\Ps_{u,v}}

\newcommand{\Kuv}{K_{u,v}}

\newcommand{\pro}{\mathbb{P}}

\newcommand{\rhoh}{\hat{\rho}}

\newcommand{\bigo}[1]{O\big(#1\big)}

\newcommand{\A}{\mathcal{A}}

\newcommand{\Thetauv}{\Theta_{u,v}}

\newcommand{\Pu}{\pro(u)}
\newcommand{\Pv}{\pro(v)}
\newcommand{\ds}{\displaystyle}

\def\Cov{\mathop{\operatorname{Cov}}}

\newcommand{\Xnk}{X_{n,k}}

\newcommand{\stdO}{\bigo{n^{ 1+(\alpha/2)\log(p)+\epsilon}}}

\newcommand{\Cuu}{C_{u,u}}
\newcommand{\Cuv}{C_{u,v}} 
\newcommand{\Cvu}{C_{v,u}}
\newcommand{\Cvv}{C_{v,v}}

\newcommand{\Ptheta}{\pro(\theta)}

\newcommand{\Psigma}{\pro(\sigma)}

\newcommand{\Pw}{\pro(w)}

\newcommand{\Qst}{\textbf{Q}_{\sigma,\theta}}
\newcommand{\deltauv}{\delta_{u,v}}

\newcommand{\Dv}{D_{v}}
\newcommand{\Du}{D_{u}}

\newcommand{\phiu}{\phi_{u}}
 \newcommand{\phiv}{\phi_{v}}

\newcommand{\RuRv}{(\Ru\Rv)}

\newcommand{\duv}{\delta_{u,v}}

\newcommand{\Ru}{R_{u}}
\newcommand{\Rv}{R_{v}}
\newcommand{\Ruv}{R_{u,v}}

\newcommand{\Qk}{\textbf{Q}_{r,c,d}^{(k)}}
\newcommand{\Tk}{\textbf{T}_{r,c,d}^{(k)}}

\newcommand{\gs}{g^{*}}

\newcommand{\Ps}{\textbf{P}}

\newcommand{\Vtt}{\widetilde{V}_{3}}

\begin{document}
\maketitle
\begin{abstract}
The precise analysis of the variance of the profile of a suffix tree
has been a longstanding open problem.
We analyze three regimes of 
the asymptotic growth of the variance of the profile of a
suffix tree built from a randomly generated binary string,
in the nonuniform case.  We utilize
combinatorics on words, singularity analysis, and the Mellin transform.
\end{abstract}

\section{Introduction}

One open problem about suffix trees is how to characterize the number
of internal nodes on the $k$th level of a suffix tree that has $n$ leaves.
Park~et al.~\cite{Park:2009} precisely analyzed the profile of
retrieval tries in 2009.  
Ward has been working on the analogous problem in suffix trees for a
decade; see, e.g.,~\cite{NicodemeWard:2011,Ward:2007}.
While the mean profile of retrieval trees and suffix trees are 
the same (asymptotically, up to first order, in the main range of
interest of the parameters), the variances of the profiles of these
two classes of trees are different.  The goal of this paper is to
precisely analyze the variance of the profile of suffix trees.

In retrieval trees, the strings inserted
into the tree structure are often considered to be independent;
such was the case in~\cite{Park:2009}.  In contrast to this, in
suffix trees, the strings inserted into the tree are suffixes of a
common string, so these strings are overlapping.  The overlaps
make the corresponding analysis much trickier, as compared to~\cite{Park:2009}.

We analyze a suffix tree built from the suffixes of a common string
$S = S_{1}S_{2}S_{3}\ldots$, where the $S_{j}$'s are randomly
generated, independent, and identically distributed.  We view each
$S_{j}$ as a letter from the alphabet $\A = \{a,b\}$, where $P(S_{j} =
a) = p$ and $P(S_{j} = b) = q$.  (Without loss of generality, we
assume throughout that $p > q$.)  
We use $\A^{\ell}$ to denote the set of words of length $\ell$.
For a word $u$ that consists of $i$ occurrences of letter
$a$ and $j$ occurrences of letter $b$, we use $\Pu$ to denote the
probability that a randomly chosen word of length $|u|$ is exactly
equal to $u$, i.e., $\Pu := p^{i}q^{j}$.

The $j$th string to be inserted into the suffix tree is
$S^{(j)} := S_{j}S_{j+1}S_{j+2}\ldots$.  We consider a randomly
generated suffix tree
$\mathcal{T}_{n}$ built over the first $n$ suffixes of $S$, i.e.,
built from the suffixes $S^{(1)}$ through $S^{(n)}$.
Briefly, all $n$ of these suffixes can be viewed as initially being
placed at the root of the suffix tree.  The $n$ suffixes are then
filtered down to the left or right children of the root, making the
classification of the suffixes according to whether the first letter
of each suffix is ``$a$'' or ``$b$'', respectively.  The filtering
continues down through the tree, with splitting at the $j$th level
according to the $j$th letter in the corresponding suffixes in that
portion of the tree.

For each word $u \in \Ak$, the suffix tree $\mathcal{T}_{n}$ will
contain the internal node corresponding to $u$ if and only if the base-string
$S$ contains at least two copies of the word $u$ within its first $n+k-1$
characters.  (Equivalently, $\mathcal{T}_{n}$ 
contains the internal node corresponding to $u$ if and only if at
least two of the suffixes $S^{(1)}$ through $S^{(n)}$ have $u$ as a prefix.)
For this reason, we define
$I_{n,u}:= 1$
if $u$ appears at least twice in $S_{1}S_{2}\ldots S_{n+k-1}$, 
or $I_{n,u}:= 0$ otherwise.
We use $X_{n,k}$ to denote the number of internal nodes in
$\mathcal{T}_{n}$ at level $k$.
With the above notation in place, we observe that 
$X_{n,k}=\sumu \Iu$.  This decomposition will be crucial to our
proofs, which start in Section~\ref{sec:combo}.

Finally, following the lead of \cite{Park:2009}, we assume that the limit $\alpha := \lim_{n \rightarrow \infty}k/\log(n)$ exists.

\section{Main Results}\label{intro}
The value of $\Var(X_{n,k})$ depends qualitatively on the quantity $\alpha,$ which describes the relationship between $n$ and $k$ via the relation $k/\log(n) \rightarrow \alpha$. It turns out that there are two particular alpha-values of importance,
\begin{align*}
\alpha_{1}=-\frac{1}{\log(q)},\hspace{30pt} \alpha_{2}=-\frac{p^{2}+q^{2}}{p^{2}\log(p)+q^{2}\log(q)}.
\end{align*}
We do not attempt, as Park et al. did in \cite{Park:2009}, to analyze
the cases where $\alpha$ is exactly equal to one of these
$\alpha_{i}$, but instead assume that both $|\alpha-\alpha_{i}|$ are
strictly positive. Given this restriction, it is permissible to take the approximation $k=\alpha\log(n)$, which we do henceforth without comment.

The variance obeys different laws depending on where the value of $\alpha$ falls in the ranges defined by these $\alpha_{i}$. 
The range of most interest is (perhaps) the range in which $\alpha_1 <
\alpha < \alpha_2$; we discuss this case in Theorem~\ref{saddlepointtheorem}.
(The case $\alpha < \alpha_1$ is discussed in 
Theorem~\ref{smallalphatheorem};
 and the case $\alpha_2<\alpha$ is handled in
Theorem~\ref{poletheorem}.)

When $\alpha$ is small, we have an easy and very strong bound on the decay of $\Var(X_{n,k})$.
\begin{theorem}\label{smallalphatheorem}
When $\alpha < \alpha_{1}$, there exists $B>0$ such that 
\begin{align*}\Var(X_{n,k})=O(e^{-n^{B}}).
\end{align*} 
\end{theorem}

The proof of Theorem \ref{smallalphatheorem} follows from lemmas that
mimic the techniques of \cite{Markthesis}; we omit it from this shortened version. The intuitive meaning behind Theorem \ref{smallalphatheorem}
is that level $k$ of the suffix tree is extremely likely to be
completely filled (meaning the variance will be extremely small) if
$\log(n)$ is sufficiently large in comparison to $k$.

Our main results deal with the less trivial case when $\alpha > \alpha_{1}$. We first introduce the functions involved in our main estimates, and provide a word on how we obtain them. 

\subsection{Functions Involved in Main Results; Methodology}\label{subsec:intro}
Our basic device for computing the variance of the internal profile is to write $X_{n,k}$ as a sum of indicator variables $\Iu$, and then evaluate
\begin{align}\label{newstarequation}
\Var(X_{n,k}) = \Var(\sumu \Iu) = \sumu \Var(\Iu) + \sumuv \Cov(\Iu,\Iv).
\end{align}

Our final analysis of the sum of the $\Var(\Iu)$ will be fairly simple: we will ultimately just have to evaluate the inverse Mellin integral
\begin{align}\label{introinversemellin}
\frac{1}{2 \pi i}\int_{c-i\infty}^{c+i\infty}n^{-s}f(s)\sum_{u \in \Ak}\Pu^{-s}\;ds=\frac{1}{2\pi i}\int_{c-i\infty}^{c+i\infty}f(s)\;n^{h(s)}\;ds,
\end{align}
where the function $h(s)$ will be given by
\begin{align*}
h(s) := -s + \alpha\log(p^{-s}+q^{-s}).
\end{align*}
(See~\cite{PF120} for more details about the Mellin transform.)
The function $h(s)$ is the same as analyzed in~\cite{Park:2009}, and their arguments extend seamlessly to our case. 

On the other hand, the terms $\Cov(\Iu,\Iv)$ for $u \neq v$  will be
novel and much more interesting. To deal with them, we will consider
all possible overlapping decompositions $(\sigma w, w\theta)$ of
$(u,v)$.  To accomplish this, we observe that 
\begin{equation}\label{bigHanalogue}
n^{-s}\sum_{\ell=1}^{k-1}\sum_{\substack{ w \in \A^{k-\ell} \\ \sigma,\theta \in \A^{\ell}}}\Pw^{-s}(\Psigma+\Ptheta)^{-s}=\sum_{\ell=1}^{k-1}\sum_{i,j=0}^{\ell}{\ell \choose i}{\ell \choose j}n^{H(s,\;(k-\ell)/k,\;i/\ell,\; j/\ell)},
\end{equation}
where $H(s,r,c,d)$ is defined as
\begin{equation*}
H(s,r,c,d) := -s+\alpha(1-r)\log(p^{-s}+q^{-s})- s\Big(\frac{\alpha}{k}\Big)\log( (p^{c}q^{1-c})^{kr} + (p^{d}q^{1-d})^{kr}).
\end{equation*}
Note: For ease of the (already cumbersome) notation, we have not written
$\alpha$ nor $k$ as a parameter of $H$.
We will substitute the right hand side of~(\ref{bigHanalogue}) for 
$n^{-s}\sum_{u\in\mathcal{A}^{k}} \Pu^{-s}$ into equation~(\ref{introinversemellin}).
We will use a technique for $H$ similar to that used for $h$, namely, summing over all
possible values $p^{i}q^{\ell-i}$ and $p^{j}q^{\ell-j}$ of $\Psigma$
and $\Ptheta$ respectively, and summing $\Pw$ into a closed form, as
was done at (\ref{introinversemellin}).

The dominant contribution to (\ref{bigHanalogue}) comes from terms
with small $r$. Since $\lim_{r \rightarrow 0}H(s,r,c,d)=h(s)$, this
implies that $\sum_{u,v}\Cov(\Iu,\Iv)$ and $\sum_{u}\Var(\Iu)$ have
the same first-order asymptotic growth, as functions of $n$.

We will evaluate the inverse Mellin integral at
(\ref{introinversemellin}) (and the analogous integral for $H$) by using either the saddle point method or by taking the residue of the pole of $\Gamma(s+2)$ at $s=-2$; which device we use will depend on the value of $\alpha$. Before giving our main results, we list the saddle points of the functions $h(s)$ and $H(s,r,c,d)$, which are
\begin{align}\label{rhodef}
\rho&:= \ds\frac{\Big(-\ds\frac{\alpha\log(p)+1}{\alpha\log(q)+1}\Big)}{\log(p/q)},\notag\\
\rho_{r,c,d}&:=\ds\frac{\Big(-\ds\frac{\alpha(1-r)\log(p)+1+(\alpha/k)\log( (p^{c}q^{1-c})^{kr} + (p^{d}q^{1-d})^{kr})}{\alpha(1-r)\log(q)+1+(\alpha/k)\log( (p^{c}q^{1-c})^{kr} + (p^{d}q^{1-d})^{kr})}\Big)}{\log(p/q)}.
\end{align}
It is also easy to verify that for any $y \in \mathbb{Z}$, the value
$s=\rho+2\pi i y/\log(p/q)$ is also a saddle point of $h$, and
similarly, $s=\rho_{r,c,d}+2\pi i y/\log(p/q)$ is a saddle point of $H$.

These saddle points will (at last) allow us to express an asymptotic value for $\Var(X_{n,k})$ in the case where $\alpha_{1} < \alpha < \alpha_{2}$.
\subsection{Behavior in the main regime}
\begin{theorem}\label{saddlepointtheorem}
Assume $\alpha$ satisfies $\alpha_{1} < \alpha <
\alpha_{2}$. Let $\rho$ and $\rho_{r,c,d}$ be as in (\ref{rhodef}). Then we have
\begin{align*}
\Var(\Xnk)=\frac{n^{h(\rho)}(C_{1}(n)+2C_{2}(n)) }{\sqrt{\log(n)}}\times \big(1+O(\log(n)^{-1})\big).
\end{align*}
%(We claim but do not prove that $\liminf_{n\rightarrow \infty}
%C_{j}(n) > 0$ and $\limsup_{n\rightarrow \infty} C_{j}(n) < \infty$.)
The $C_{1}(n)$ is given by
\begin{align*}
C_{1}(n)=\sum_{y \in \mathbb{Z}}\frac{n^{i \Im(h(\rho+iyK))}f_{1}(\rho+iyK)\Gamma(\rho+iyK+1)}{\sqrt{2\pi h''(\rho+iyK)}},
\end{align*}
where $K:=2\pi/\log(p/q)$ and where $f_{1}(s):=1-2^{-s}-s2^{-s-2}$.
Regarding $C_{2}(n)$, we define
%if we use the shorthand notation 
%\begin{align*}
$r=\frac{\ell}{k}$, $c=\frac{i}{\ell}$, $d=\frac{j}{\ell}$,
%\end{align*}
and then $C_{2}(n)$ is given by
\begin{align*}
C_{2}(n)&=\sum_{\substack{ 0 < \ell < k \\ 0 \leq i,j \leq \ell}}{\ell \choose i}{\ell \choose j}\frac{n^{H(\rho_{r,c,d},r,c,d)}}{n^{h(\rho)}}\sum_{y \in \mathbb{Z}}\frac{n^{i \Im(H(\rho_{r,c,d}+iyK,r,c,d))}f_{2}(\rho_{r,c,d}+iyK,\ell,i,j)\Gamma(\rho_{r,c,d}+iyK+2)}{\sqrt{2\pi \frac{\partial H}{\partial s}(\rho_{r,c,d}+iyK,r,c,d)}}\\
& \times (1+O(\log(n)^{-1})).
\end{align*}
with the function $f_{2}(s,\ell,i,j)$ given by
\begin{align*}
f_{2}(s,\ell,i,j)=\sum_{m \geq 2} \Big(\frac{p^{i}q^{\ell-i}p^{j}q^{\ell-j}}{p^{i}q^{\ell-i}+p^{j}q^{\ell-j}}\Big)^{m-1}\frac{\Gamma(s+m)}{\Gamma(s+2) m!} L_{m}\Big(\frac{p^{i}q^{\ell-i}p^{j}q^{\ell-j}}{p^{i}q^{\ell-i}+p^{j}q^{\ell-j}}, \;\frac{p^{i}q^{\ell-i}p^{j}q^{\ell-j}}{(p^{i}q^{\ell-i}+p^{j}q^{\ell-j})^{2}},\;s+m\Big),
\end{align*}
with
\begin{align*}
L_{m}(a,b,x)=a(m-1)^{2} +m(2-m)+bmx.
\end{align*}
Furthermore, the outer sum in $C_{2}(n)$ satisfies the decay condition that for any positive integer $\ell_{0}$, the sum over all $\ell>\ell_{0}$ and $1 \leq i,j \leq \ell$ is $O(n^{-(\ell_{0}/k) \times \beta})$ for a fixed $\beta>0$.

\end{theorem}

\subsection{Behavior in the polar regime}
In the final $\alpha$-regime, where $\alpha> \alpha_{2}$, the
asymptotics arise from the pole at $s=-2$, as the following theorem states.
\begin{theorem}\label{poletheorem}
Assume the parameter $\alpha$ satisfies $\alpha>\alpha_{2}$. Then for some $\epsilon>0$, we have
\begin{align*}
\Var(X_{n,k})=n^{h(-2)}(C_{1}(n)+2C_{2}(n)) \times (1+O(n^{-\epsilon}))
\end{align*}
with $f_{1},f_{2}$ as defined in Theorem
\ref{saddlepointtheorem}, and $C_{1}(n)$, $C_{2}(n)$ are given by
\begin{align*}
C_{1}(n)=f_{1}(-2),\hspace{20pt} C_{2}(n)=f_{2}(-2)\sum_{\substack{ 0 < \ell < k \\ 0 \leq i,j \leq \ell}}{\ell \choose i}{\ell \choose j}\frac{n^{H(-2,r,c,d)}}{n^{h(-2)}}
\end{align*}
with the decay of $C_{2}(n)$ as in Theorem \ref{saddlepointtheorem}.
\end{theorem}

Having stated our main results, we now proceed to the proof of Theorems \ref{saddlepointtheorem} and \ref{poletheorem}, which will occupy the remainder of the paper.

\section{An Expression for the Variance}\label{sec:combo} 
Our first task in proving  Theorems \ref{saddlepointtheorem} and
\ref{poletheorem} is  to obtain an exact expression for the variance
of the internal profile $X_{n,k}$.
Recalling equation~(\ref{newstarequation}),
%~(\ref{XfromI}),
%we have $X_{n,k}=\sumu \Iu$, and therefore
%$$\Var(\X) =\sumu\Var(\Iu) + \sumuv \Cov(\Iu,\Iv).$$
we need to derive the values of $\Var(\Iu)$ and $\Cov(\Iu,\Iv)$, so we let
$U_{n}$ denote the number of occurrences of $u$ in the first $n$
characters of $S$,
and we define $V_{n}$ analogously. Then inclusions-exclusion yields the representations
\begin{align}\label{indicatorvar}
\Var(\Iu)&=\Big(1-\sum_{i=0}^{1}\pro(U_{n+k-1}=i)\Big)-\Big(1-\sum_{i=0}^{1}\pro(U_{n+k-1}=i)\Big)^{2}\notag\\
\Cov(\Iu,\Iv)&=\sum_{0 \leq i,j \leq 1}\Big(\pro(U_{n+k-1}=i,\;V_{n+k-1}=j)-\pro(U_{n+k-1}=i) \times \pro(V_{n+k-1}=j)\Big)
\end{align}
where we require $u$ and $v$ to be distinct. Thus, to obtain an expression for $\Var(X_{n,k})$, we just have to evaluate all the probabilities in (\ref{indicatorvar}).

\section{Explicit Expressions for Word-Occurrence Probabilities}\label{sec:Cauchy}
To estimate the probabilities in (\ref{indicatorvar}), we use
generating functions, 
%which are the standard tool employed in this
%context. As an illustrative example, to determine $\pro(U_{n+k-1}=0)$,
%we will first construct an explicit expression for the power series
%\begin{align}\label{genericgeneratingfunction}
%\sum_{n \geq 0} z^{n}\pro(U_{n+k-1}=0),
%\end{align}
and %then use 
complex analysis.
% to estimate the coefficient of $z^{n}$.
%
%To construct generating functions of the form of
%(\ref{genericgeneratingfunction}), w
Motivated by~\cite{Bassino:2012}, 
%which gives a general form for such functions.
%
%To define our probability generating functions of form
%(\ref{genericgeneratingfunction}), we first introduce the
%foundational polynomials 
we define
\begin{equation}\label{psidef}
\psi(z)=\Cuu(z)\Cvv(z)-\Cuv(z)\Cvu(z),\qquad\hbox{and}\qquad
\phiu(z)=\Cvv(z)-\Cuv(z),%\; \phiv(z)=\Cuu(z)-\Cvu(z),
\end{equation}
where the functions $C_{x,y}(z)$ are \emph{correlation polynomials},
the fundamental device for dealing with the phenomenon of
word-overlaps. With these functions in hand, we can define
generating-functions 
%of form (\ref{genericgeneratingfunction}) 
for all the probabilities in (\ref{indicatorvar}). We summarize the result in the following proposition.
\begin{proposition}
Let $\psi(z)$ and $\phi_{u}(z)$ be as defined at (\ref{psidef}), and define the functions 
\begin{align}\label{deltadef}
D_{u}(z)&=(1-z)\Cuu(z)+ z^{k}\Pu,\;\;\qquad\delta_{u,v}(z)=(1-z)\psi(z) + z^{k}(\phiu(z)\Pu + \phiv(z)\Pv),\cr
G_{0}^{(u)}(z)&=C_{u,u}(z),\ \ G_{1}^{(u)}(z)=\Pu z^{k}, 
G_{0,0}^{(u,v)}(z)=\psi(z),\ \ 
G_{1,0}^{(u,v)}(z)=\delta_{u,v}(z)\Cvv(z)-\psi(z)\Dv(z),\cr
G_{1,1}^{(u,v)}(z)&=\delta_{u,v}(z)^{2} -\delta_{u,v}(z)\big(\Cvv(z)\Du(z)+\Cuu(z)\Dv(z)+(1-z)\psi(z)\big)+ 2\psi(z)\Du(z)\Dv(z),
\end{align}
with all $v$-counting functions defined in a manner analogous to the $u$-counting functions. Then we have the closed-form power series expressions
\begin{align}\label{Gudef}
\frac{G_{i}^{(u)}(z)}{D_{u}(z)^{i+1}}=\sum_{n \geq
  0}z^{n}\pro(U_{n}=i),\quad\hbox{and}\quad
\frac{G_{i,j}^{(u,v)}(z)}{\delta_{u,v}(z)^{i+j+1}}=\sum_{n \geq 0}z^{n}\pro(U_{n}=i,\;V_{n}=j),\;\;\;\; 0 \leq i,j \leq 1.
\end{align}
\end{proposition}

Now we must derive the $(n+k-1)$st coefficients of these generating
functions. To do this, we use Cauchy's Integral Formula, following a standard argument in combinatorics on words.  Our specific methodology will rely on a vital fact about the denominators $D_{u}(z),D_{v}(z)$ and $\deltauv(z)$ of the probability generating functions in (\ref{Gudef}).

\begin{lemma}\label{uniquerootlemma}
There exist $K,\rho>0$ such that for all $k > K$ and all $u, v \in \Ak$, 
each of the polynomials $D_{u}(z)$, $D_{v}(z)$, and $\delta_{u,v}(z)$ has a unique root (defined respectively as $\Ru,\Rv$ and $\Ruv$) in the disc $|z| \leq \rho$.
\end{lemma}
The proof for $D_{u}(z)$ and $D_{v}(z)$ is given in \cite{Jacquet:2005}; spatial constraints prevent us from giving the proof for the $\delta_{u,v}(z)$ portion.

Armed with Lemma \ref{uniquerootlemma}, we can estimate the word-counting coefficients of our generating functions to within a factor of $O(\rho^{-n})$ by applying Cauchy's Theorem to the contour $z=|\rho|$. The following theorem gives the resultant estimates.
\begin{theorem}\label{restheorem}
Let the polynomials $D_{u},D_{v},\deltauv$ and $G_{0}^{(u)},
G_{1}^{(u)}$, etc.\ be as in (\ref{deltadef}) and
(\ref{Gudef}).
If we define
\begin{align*}
c_{0,0}^{(u)}=-\frac{C_{u,u}(\Ru)}{D_{u}'(\Ru)},\;\;\;\;\;c_{1,0}^{(u)}=\frac{\Pu D_{u}''(\Ru)}{D_{u}'(\Ru)^{3}},\;\;\;\;\; c_{1,1}^{(u)}=\frac{\Pu}{D_{u}'(\Ru)^{2}},
\end{align*}
then we have the following estimates
\begin{equation*} 
\pro(U_{n+k-1}=0)\approx c_{0,0}^{(u)}\frac{1}{\Ru^{n+k}},\qquad\hbox{and}\qquad
\pro(U_{n+k-1}=1)\approx c_{1,0}^{(u)}\frac{1}{\Ru^{n}}+c_{1,1}^{(u)}\frac{n}{\Ru^{n+1}},
\end{equation*}
and the error in each case is $O(\rho^{-n})$.

Similarly, for the joint events $(U_{n+k-1}=i,\;V_{n+k-1}=j)$,
and %the constants $a_{i,j}^{(u,v)}$ describing both $U_{n+k-1}$ and $V_{n+k-1}$ are
\begin{align*}
a_{0,0}^{(u,v)}&=-\frac{\psi'(\Ruv)}{\duv'(\Ruv)},\;\;\;\;\;a_{1,0,u}^{(u,v)}=-\frac{G_{1,0}^{(u,v)}(\Ruv) \duv''(\Ruv)}{\delta'(\Ruv)^{3}},\;\;\;\;\;a_{1,1,u}^{(u,v)}=\frac{G_{1,0}^{(u,v)}(\Ruv)}{\delta'(\Ruv)^{2}},\\
a_{2,0}^{(u,v)}&=-\frac{G_{1,1}^{(u,v)}{}''(\Ruv)}{2\duv'(\Ruv)^{3}}+\frac{3G_{1,1}^{(u,v)}{}'(\Ruv)\duv{}''(\Ruv)}{2\delta_{u,v}'(\Ruv)^{4}}\\
&\qquad{}-\frac{G_{1,1}^{(u,v)}(\Ruv)(-\delta_{u,v}'(\Ruv)\delta_{u,v}'''(\Ruv)+3\delta_{u,v}''(\Ruv)^{2})}{2\delta_{u,v}'(\Ruv)^{5}},\\
a_{2,1}^{(u,v)}&=\frac{G_{1,1}^{(u,v)}{}'(\Ruv)}{\delta_{u,v}{}'(\Ruv)^{3}}-\frac{3G_{0,0}^{(u,v)}(\Ruv)\delta_{u,v}{}''(\Ruv)}{2\delta_{u,v}{}'(\Ruv)^{4}},\;\;\;\;\; a_{2,2}^{(u,v)}=-\frac{G_{1,1}^{(u,v)}(\Ruv)}{2\delta_{u,v}'(\Ruv)^{3}},
\end{align*}
with $G_{i,j}^{(u,v)}(z)$ as in (\ref{Gudef}), we also obtain these
estimates, where again, the error in each case is $O(\rho^{-n})$:
\begin{align*}
\pro(U_{n+k-1}=0,V_{n+k-1}=0)&\approx a_{0,0}^{(u,v)}\frac{1}{\Ruv^{n+k}},\\
\pro(U_{n+k-1}=1,V_{n+k-1}=0)&\approx a_{1,0,u}^{(u,v)}\frac{1}{\Ruv^{n+k}}+a_{1,1,u}^{(u,v)}+\frac{(n+k)}{\Ruv^{n+k+1}},\\
\pro(U_{n+k-1}=1,V_{n+k-1}=1)&\approx a_{2,0}^{(u,v)}\frac{1}{\Ruv^{n+k}}+a_{2,1}^{(u,v)}\frac{(n+k)}{\Ruv^{n+k+1}}+a_{2,2}^{(u,v)}\frac{(n+k)(n+k+1)}{\Ruv^{n+k+2}}.
\end{align*}
\end{theorem}
Using these expressions, we can evaluate the expressions for
$\Var(\Iu)$ and $\Cov(\Iu,\Iv)$ at (\ref{indicatorvar}) to within a
factor of $O(\rho^{-n})$. In doing this, however, it will be helpful
to break up our estimates from Theorem \ref{restheorem} so that terms
of common order in $n$ are denoted under a single variable. We
therefore define the upper-case constants (we suppress the dependence
on $u$ and $v$ in the notation)

\begin{align}
C_{0}&=\frac{c_{0,0}^{(u)}+c_{1,0}^{(u)}}{\Ru^{k}} + \frac{k
  c_{1,1}^{(u)}}{\Ru^{k+1}},\;\;\;\;\;C_{1}=\frac{c_{1,1}^{(u)}}{\Ru^{k+1}},
\notag\\
A_{0} &= \frac{a_{0,0}^{(u,v)}+a_{1,0,u}^{(u,v)}+a_{1,0,v}^{(u,v)}+a_{2,0}^{(u,v)}}{\Ruv^{k}}+\frac{\big(a_{1,1,u}^{(u,v)}+a_{1,1,v}^{(u,v)}\big)k}{\Ruv^{k+1}}+\frac{a_{1,1}^{(u,v)}k(k+1)}{\Ruv^{k+2}},\notag\\
A_{1} &= \frac{a_{1,1,u}^{(u,v)}+a_{1,1,v}^{(u,v)}+a_{2,1}^{(u,v)}}{\Ruv^{k+1}}+\frac{a_{2,2}^{(u,v)}(2k+1)}{\Ruv^{k+2}},\;\;\;\;\; A_{2}=\frac{a_{2,2}}{\Ruv^{k+2}},
\qquad B_{0}=\frac{c_{0,0}^{(v)}c_{0,0}^{(u)}}{\RuRv^{k}},\;\;\;\;\;\notag\\
B_{1}&=\Big(c_{1,0}^{(u)}+\frac{c_{1,1}^{(u)}}{\Ru}\Big)\frac{c_{0,0}^{(u)}}{\Rv^{k}}+\Big(c_{1,0}^{(v)}+\frac{c_{1,1}^{(v)}}{\Rv}\Big)\frac{c_{0,0}^{(u)}}{\Ru^{k}},
\qquad B_{2}= \Big(c_{1,0}^{(u)}+\frac{c_{1,1}^{(u)}}{\Ru}\Big)\Big(c_{1,0}^{(v)}+\frac{c_{1,1}^{(v)}}{\Rv}\Big).\label{Adef}
\end{align}
%The virtue of these abbreviated expressions is that they allow us to
%never have to mention $a_{i,j}^{(u,v)}$ and its disgusting cousins
%ever again. Now, plugging into
Returning to the expression $\Var(X_{n,k})=\sumu \Var(\Iu) + \sumuv \Cov(\Iv)$, we obtain an expression for our ultimate desired quantity.
\begin{corollary}\label{Acorr}
Let $A_{i},B_{i},C_{i}$ be as defined in (\ref{Adef}). 
With $A_{i},B_{i}$ and $C_{i}$ as in (\ref{Adef}),
we have the estimate
\begin{equation*}
\Var(\Xnk)= \sumu \Big(1-\frac{C_{0}+nC_{1}}{\Ru^{n}}\Big)- \Big(1-\frac{C_{0}+nC_{1}}{\Ru^{n}}\Big)^{2}
+\sumuv\sum_{i=0}^{2} \Big(\frac{A_{i}}{\Ruv^{n}} - \frac{ B_{i}}{\RuRv^{n}}\Big)n^{i} + O(\rho^{-n}).
\end{equation*}
\end{corollary}

\subsection{High-Probability Approximations}
Our task is now to approximate the expression from Corollary~\ref{Acorr}. To achieve this, we follow the usual suffix-tree strategy: we compare the terms to simpler ones which will be accurate with very high probability, and use Mellin transforms to show that sum of the the differences between the old terms and the new ones is negligible. Our two main tools for demonstrating this negligibility are bounds provided by the following lemma.

%We construct our estimates using the heuristic ``With very high probability, $\Cuv(1) \approx 1$ and $\Cuv(1)\Cvu(1) \approx 0$''. This heuristic is formalized in the following Lemma.
\begin{lemma}\label{corrlemma}
We have the bounds
\begin{align*}
\sum_{u \in \Ak} \Pu (\Cuu(1)-1) &=O(p^{k/2}),\hspace{23pt}
\sumuv \Pu\Cuv(1)\Cvu(1)= O(p^{k/2})
\end{align*}
\end{lemma}
The first portion of Lemma \ref{corrlemma} is proved in
\cite{Jacquet:2005}; spatial constraints prevent us from proving the
second portion here. However, by rigorously expanding on the heuristic $\Cuu(1) \approx 1$ and $\Cuv(1)\Cvu(1) \approx 0$, we obtain the following theorem which is one of the major steps of the proof.
\begin{theorem}\label{mellintheorem}
We define the terms
%\begin{equation*}
$\Puv := \Pu+\Pv,$,
$\Thetauv := \Pu\Cuv(1)+\Pv\Cvu(1)$,
and $\Kuv = (2k-1)\Pu\Pv$,
and the expressions

\begin{align*}
V_{1}(n)&:=\sumu 1- (1+n\Pu)e^{-n\Pu}- \big(1- (1+n\Pu)e^{-n\Pu}\big)^{2},\\  
V_{2}(n)&:=\sumuv n^{3}\Pu\Pv\Kuv e^{-n(\Puv-\Thetauv)},\\
V_{3}(n)&:=\sumuv e^{-n\Puv}(e^{n\Thetauv}-1)\big(1 + n\Puv + n^{2}\Pu\Pv)- e^{-n(\Puv-\Thetauv)}n\Thetauv\big(1+n(\Puv-\Thetauv)\big).
\end{align*}
Then, for every $\epsilon>0$, we have the estimate
\begin{align*}
\Var(X_{n,k})&= V_{1}(n)-V_{2}(n)+ 2V_{3}(n)+\stdO.
\end{align*}
\end{theorem}
We mention that the term $V_{1}(n)$ has already been analyzed in Park~\cite{Park:2009}.
It gives the asymptotic variance of the internal profile in a
\emph{trie}.
The term $V_{2}(n)$ is negligible. Thus, after proving Theorem \ref{mellintheorem}, all that will remain will be to analyze $V_{3}(n)$.

\section{Distilling Essence of Estimate}
We must now analyze the estimate from Theorem \ref{mellintheorem}, which consists of the terms $V_{1}(n)$, $V_{2}(n)$ and $V_{3}(n)$. We can deal with the first two of these terms in two quick theorems.
Theorem~\ref{Parktheorem} was proven in \cite{Park:2009}.
Theorem \ref{Ktheorem} has a short proof, which we omit  in this concise version.
\begin{theorem}\label{Parktheorem}
An asymptotic expression for $V_{1}(n)$ is given by the $C_{1}(n)$ portions from Theorems \ref{saddlepointtheorem} and \ref{poletheorem}.
\end{theorem} 
\begin{theorem}\label{Ktheorem}
The term $V_{2}(n)$ from Theorem \ref{thetatheorem} satisfies
$V_{2}(n)=\Var(X_{n,k})O(n^{-\epsilon})$
for some $\epsilon>0$.
\end{theorem}

For the rest of the paper, then, we concentrate on the portion $V_{3}(n)$, which contains the term $\Thetauv=
\Pu\Cuv(1)+\Pv\Cvu(1)$ and constitutes the really novel part of the whole enterprise. We deal with $\Thetauv$ by nothing that, by Lemma \ref{corrlemma}, the quantities $\Cuv(1)$ and $\Cvu(1)$ are unlikely
to simultaneously be large, so the approximation
$\Thetauv \approx \Pu\Cuv(1)$ is reasonable. From here, we note that for $\Thetauv$ to be nonzero we must have $\Cuv(1)>0$, in which case there exists some maximal suffix of $u$ which is also a
prefix of $v$. If we call this word $w$, and then have the precise
equality $\Pu\Cuv(1)=\Psigma \Pw \Ptheta C_{w,w}(1).$ where
$\sigma,\theta \in \A^{k-|w|}$ are such that $u=\sigma w$ and
$v=w\theta$. Then we employ the estimate $C_{w,w}(1) \approx 1$, again as suggested by Lemma \ref{corrlemma}. We thus have the central estimate
$\Thetauv \approx \Psigma \Pw \Ptheta$.
Our strategy, then, is to make the substitutions $u=\sigma w$,
$v=w\theta$, and $\Thetauv = \Psigma\Pw\Ptheta$ in the summand of $V_{3}(n)$, 
and then sum over all possible such decompositions. In the proof and final result it will be  helpful to have the shorthand 
$\Qst := \Psigma + \Ptheta$ and $\Tst := \Psigma\Ptheta$,
The following theorem states that this heuristic can be rigorously justified.
\begin{theorem}\label{thetatheorem}
Let $\Qst,\Tst$ be as defined above, and define the functions
\begin{align*}
\gt(n)&=e^{-n\Pw\Qst}(e^{x\Pw\Tst}-1)\big(1 + x\Pw\Qst + n^{2}\Pw^{2}\Tst)\\
&\qquad{} - e^{-x\Pw(\Qst-\Tst)}x\Pw\Tst\big(1+x\Pw(\Qst-\Tst)\big)
\end{align*}
and
$\Vtt(n):=\sum_{\ell=1}^{k-1}\sum_{\substack{w \in \A^{k-\ell} \\ \sigma,\theta \in \A^{\ell}}}\gt(n)$.
Then for $V_{3}(n)$ as given in Theorem \ref{mellintheorem}, we have the estimate
\begin{align*}
V_{3}(n) = 2\Vtt(n) + \stdO.
\end{align*}
\end{theorem}
One proves Theorem \ref{thetatheorem} by making the substitutions $\Pw \Qst \approx \Puv$ and $\Pw\Tst \approx \Thetauv$, and then using Mellin transforms and Lemma \ref{corrlemma} to show that the derived error-bound is satisfied.

\section{Derivation of Asymptotics}
To complete the main proof, it remains only to analyze $\Vtt(n)$. We present the key results in this process in a series of subsections.
\subsection{Partitioning the Sum}
Our first step is to partition the sum which comprises $\Vtt(n)$. 
into subsets which share a common value for the ordered pair $(\Psigma,\Ptheta)$. We can rewrite the function $\gt(n)$ from Thereom \ref{thetatheorem} as an infinite sum, 
\begin{align*}
\gt(x)&=e^{-x\Pw\Qst}\sum_{ m \geq 2}\frac{(x\Pw)^{m}\Tst^{m-1}\Qst}{m!}L_{m}\Big(\frac{\Tst}{\Qst}, \;\frac{\Pw\Tst}{\Qst},\;x\Big).
\end{align*}
with the function $L_{m}$  given by
$L_{m}(a,b,x):=a(m-1)^{2} +m(2-m)+bmx$.
The terms $\Qst$ and $\Tst$ only depend on the \emph{probabilities} of $\sigma$ and $\theta$; their internal composition does not matter. This allows a great reduction in the number of terms to handle. With some abuse of notation, we define the terms
\begin{align*}
\textbf{Q}_{r,c,d}^{(k)} &:= \textbf{Q}_{a^{krc}b^{kr(1-c)},a^{krd}b^{kr(1-d)}} = p^{krc}q^{kr(1-c)}+p^{krd}q^{kr(1-d)},\\
\Tk&:=\textbf{T}_{a^{krc}b^{kr(1-c)},a^{krd}b^{kr(1-d)}}=p^{krc}q^{kr(1-c)} \times p^{krd}q^{kr(1-d)}
\end{align*}
and then define the atom of all our remaining analysis, which is
\begin{align}\label{gdef}
g(x,r,c,d)&=\sum_{w \in \A^{k(1-r)}}e^{-x\Pw\Qk}\sum_{ m \geq 2}\frac{(x\Pw)^{m}\Tk{}^{m-1}\Qk}{m!}L_{m}\Big(\frac{\Tk}{\Qk}, \;\frac{\Pw\Tk}{\Qk},\;x\Big).
\end{align}
With this notation, we have the following proposition.
\begin{proposition}\label{gprop}
Let $g(x,r,c,d)$ be as in (\ref{gdef}). Then $\Vtt(n)$ from Theorem \ref{thetatheorem} admits the representation \begin{align}\label{hsum}
\Vtt(n)=\sum_{\substack{0 < \ell < k \\ 0 \leq i,j \leq \ell}} {\ell \choose i}{\ell \choose j}g(n,\tfrac{\ell}{k},\tfrac{i}{\ell},\tfrac{j}{\ell}).
\end{align}
\end{proposition}
Now we analyze $g$.
\subsection{Analysis of $g(n,r,c,d)$}
All our final estimates rest on our analysis of the function $g$ given in Proposition \ref{gprop}. To begin that analysis, we take the Mellin transform of $g$ and, specifying the bounded portion
\begin{align*}
W(s,r,c,d)=\sum_{m \geq 2} \Big(\frac{\Tk}{\Qk}\Big)^{m-1}\frac{\Gamma(s+m)}{\Gamma(s+2) m!} L_{m}\Big(\frac{\Tk}{\Qk}, \;\frac{\Tk}{\Qk{}^{2}},\;s+m\Big),
\end{align*}
we obtain
\begin{align*}
\gs(s,r,c,d)&=\Gamma(s+2)W(s,r,c,d)\Qk{}^{-s}\sum_{w \in \A^{k(1-r)}}\Pw^{-s}\\
&=\Gamma(s+2)W(s,r,c,d)\Qk{}^{-s}(p^{-s}+q^{-s})^{k(1-r)}.
\end{align*}

We then consider the value of $n^{-s}\gs(s,r,c,d)$, which will be the integrand of our inverse Mellin integral. Using the relation $k=\alpha\log(n)$, we can write
$n^{-s}\gs(s,r,c,d)=\Gamma(s+2)W(s,r,c,d)n^{H(s,r,c,d)}$,
where the function $H$ is as defined in Section \ref{subsec:intro}.
From here, we can recover the value of $g(n,r,c,d)$ via an inverse Mellin transform. We summarize the results in the following theorem.
\begin{theorem}\label{hasymptote}
Define the discriminant
\begin{align*}
A(r,c,d)=\frac{\alpha(1-r)}{(\alpha/k)\log(\Qk)+1}.
\end{align*}
Then the function $g(n,r,c,d)$ defined in (\ref{gdef}) obeys the following asymptotic scheme. \\
If $A(r,c,d) < \alpha_{1}$, then $g(n,r,c,d)=O(n^{-M})$ for every $M>0$.\\
If $\alpha_{1} < A(r,c,d) < \alpha_{2}$, then
\begin{align*}
g(n,r,c,d) &= \frac{n^{H(\rho_{r,c,d},r,c,d)}}{\sqrt{\log(n)}}\sum_{y \in \mathbb{Z}}\frac{n^{i \Im(H(\rho_{r,c,d}+iyK,r,c,d))}W(\rho_{r,c,d}+iyK,r,c,d)\Gamma(\rho_{r,c,d}+iyK+2)}{\sqrt{2\pi \frac{\partial H}{\partial s}(\rho_{r,c,d}+iyK,r,c,d)}}\\
&\times (1+O(\log(n)^{-1/2})).
\end{align*}
If $A(r,c,d) > \alpha_{2}$, then 
$g(n,r,c,d)=n^{H(-2,r,c,d)}W(-2,r,c,d)(1+O(n^{-\epsilon}))$
for some $\epsilon>0$.
\end{theorem}
The estimates of Theorem \ref{hasymptote} can be derived using techniques that are standard (albeit pretty technical) in the analysis of tree structures. In the first regime, one can show that $H(s,r,c,d)$ is always decreasing in $s$, so integrating along $\Re(s)=s_{0}$ for $H(s_{0})=-M$ gives the desired bound. In the second regime we use the saddle-point method, and in the final regime, we derive the asymptotics by taking the residue from the pole of $\Gamma(s+2)$ at $s=-2$.

Theorem \ref{hasymptote}, though certainly essential, is not in itself sufficient for our purposes, since we have to sum $g(n,\frac{\ell}{k},\frac{i}{\ell},\frac{j}{\ell})$ over a set of triplets $(\ell,i,j)$ that will grow unboundedly large as $n \rightarrow \infty$. The next lemma gives the needed statement about uniform convergence.
\begin{lemma}\label{uniformlemma}
Suppose $\alpha_{1} < \alpha < \alpha_{2}$. Then there exists $r_{0} > 0$ such that for all triplets $(r,c,d)$ in the rectangle $R_{0}=[0,r_{0}]\times [0,1]^{2}$, we have $\alpha_{1} < A(r,c,d) < \alpha_{2}$, and the saddle-point estimate of Theorem~\ref{hasymptote} holds uniformly. Furthermore, the analogous result holds in the polar case, when $\alpha > \alpha_{2}$.
\end{lemma}
The claims about $A(r,c,d)$ lying in particular ranges follow easily from the definition of $A(r,c,d)$. To show uniformity in the saddle point case, we use bounds from~\cite{Olver}, which are uniform on the compact set $R_{0}$. In the polar regime, we again use the compactness of $R_{0}$ to show that the $s$-partial of $H(s,r,c,d)$ at $s=0$ is bounded below by a positive constant, meaning that for some $\epsilon>0$, we can uniformly take the left-hand side our Mellin box to be $\Re(s)=-2-\epsilon$, thereby obtaining an error that is $O(n^{H(-2-\epsilon,r,c,d)})$, with the $(r,c,d)$ portion controlled by compactness.

\section{Bounding the Tail}\label{tailsection}
Theorem \ref{hasymptote} justifies the content of $C_{2}(n)$ in the main Theorems \ref{saddlepointtheorem} and \ref{poletheorem}. However, we still have to justify the uniform $(1+O(\cdot))$ error-bounds given in the leading equations of those theorems (which amounts to showing that our estimates for $g(n,r,c,d)$ are uniform outside the compact rectangle $R_{0})$ as well as prove our claim about the decay of the outer sum in $C_{2}(n)$. 

We can accomplish both these tasks using the same argument. First, we unify the $s$-arguments for $H$ in the polar and saddle-point cases into a single term,
\begin{align}\label{rhohdef}
\rhoh_{r,c,d} := \begin{cases} \rho_{r,c,d} &: \alpha_{1} < \alpha < \alpha_{2}\\
-1 &: \alpha > \alpha_{2}.\end{cases}
\end{align}
Then we note that if we define
\begin{align}\label{Gdef}
G(r,c,d)=\alpha r (-c\log(c)-(1-c)\log(1-c)-d\log(d)-(1-d)\log(1-d)) + H(\rhoh_{r,c,d},r,c,d),
\end{align}
then by Stirling's Formula we have 
\begin{align*}
{kr \choose krc}{kr \choose krd}g(n,r,c,d)=n^{G(r,c,d)} \times Y(\log(n)),
\end{align*}
where the function $Y(\log(n))$ is unimportant except for the fact that its growth/decay are in $\log(n)$. We now state an important and somewhat surprising result about the function $G$.
\begin{lemma}\label{maxlemma}
Let the function $G(r,c,d)$ be as in (\ref{Gdef}), and  $A(r,c,d)$ the discriminant from Theorem~\ref{hasymptote}. Then for any fixed $r$ such that the set\; $\Omega_{r}:=\{(c,d)\;:\;A(r,c,d)>\alpha_{1}\}$ is nonempty, the map $(c,d) \rightarrow G(r,c,d)$ attains its maximum at a unique ordered pair $(c_{m}(r),c_{m}(r))$ on the diagonal of $\Omega_{r}$. 
\end{lemma}
The proof of Lemma \ref{maxlemma}, although not exceedingly difficult or technical, is rather long and (to us) not very intuitive. We therefore omit it.
Lemma \ref{maxlemma} allows us to define the function
\begin{align}\label{Fdef}
F(r)=G(r,c_{m}(r),c_{m}(r))
\end{align}
for every $r$ on which the set $\Omega_{r}$ defined in Lemma \ref{maxlemma} is nonempty. We now state two vital facts about this $F$, which are exactly the results needed complete the proof.
\begin{lemma}\label{Flemma}
The function $F(r)$ defined at (\ref{Fdef}) is concave, and moreover $lim_{r \rightarrow 0}F'(r)<0$.
\end{lemma}
The statements in Theorems \ref{saddlepointtheorem} and \ref{poletheorem} about the decay of $C_{2}(n)$ immediately follow from Lemma \ref{Flemma}, since we have
$n^{F(0)-(\ell/k)F'(0)} \geq n^{F(\ell/k)} \geq {\ell \choose i}{\ell \choose j}n^{H(\rhoh_{r,c,d},r,c,d)}$,
and one readily verifies that $F(0)=h(\rho)$ in the saddle-point case and $h(0)$ in the polar case.
It remains only to justify the global $O$-bounds at the beginning of Theorems \ref{saddlepointtheorem} and \ref{poletheorem} for those $(r,c,d)$ outside the rectangle $R_{0}$ given in Lemma \ref{uniformlemma}, which the following achieves.
\begin{lemma}\label{uniformboundlemma}
With $F$ as defined at (\ref{Fdef}) and $g$ as at (\ref{gdef}), for all sufficiently small $r_{0}$ there exists $C$ such that
\begin{align*}
{kr \choose krc}{kr \choose krd}g(n,r,c,d) \leq C n^{F(0)-(r_{0}/2)F'(0)}
\end{align*}
for all $r>r_{0}$ and all $(c,d) \in [0,1]$.
\end{lemma}
The main tool in proving Lemma \ref{uniformboundlemma} is Lemma \ref{Flemma}, although some work is required in proving uniformity in (for example) cases where the saddle point $\rhoh_{r,c,d}$ is very close to the pole at $s=-2$.

\nocite{*}
\acknowledgements
\label{sec:ack}
Ward thanks the authors of~\cite{Bassino:2012} for their hospitality
during several visits to Paris, which enabled him to work on this
problem.  Both authors also thank the authors of~\cite{Park:2009} for
their pioneering work in this area.  Moreover, we both acknowledge
W.~Szpankowski as a constant source of encouragement during the years
in which we worked toward a method of solution.
Finally, we thank the three anonymous referees for their very
insightful comments, corrections, and suggestions.

%\bibliographystyle{abbrvnat}
% use the following instead if you encounter problems 
\bibliographystyle{alpha}
\bibliography{gaitherward}
\label{sec:biblio}

\end{document}